\documentclass[a4paper,11pt]{article}

\usepackage{jheppub} 
\usepackage{bm,latexsym,amsmath,amsfonts,amssymb,fancyhdr,color,graphicx,slashed}
\usepackage[usenames,dvipsnames,svgnames,table]{xcolor}
\usepackage[normalem]{ulem}
\usepackage{soul}
\usepackage{fontawesome5} 

\setstcolor{red}

\makeatletter
\global\def\@fpheader{\,}
\makeatother

\graphicspath{{figs/}}

\title{Dark Matter Attenuation inside the Earth: A Boltzmann Equation Approach}

\author[a]{Chuan-Yang Xing,}
\affiliation[a]{College of Science, China University of Petroleum (East China), Qingdao 266580, China}
\emailAdd{cyxing@upc.edu.cn}

\author[b,c]{Chen Xia}
\affiliation[b]{Shanghai Synchrotron Radiation Facility, Shanghai Advanced Research Institute, Chinese Academy of Sciences, Shanghai 201204, China}
\affiliation[c]{Shanghai Key Laboratory of Particle Accelerator Science and Technology, Shanghai 201210, China}
\emailAdd{xiac@sari.ac.cn}

\abstract{
For strongly interacting or boosted dark matter, propagation through the Earth can involve sizable scattering and energy loss, reshaping the underground flux in energy, direction, and normalization. Scattered particles may still fall within the detector acceptance, so the detector-side signal depends on phase-space transport from the Earth's surface to the underground detector. In this work, we formulate this transport problem with the Boltzmann equation. Its integral solution organizes successive scattering effects as a deterministic expansion in scattering orders. We analyze the transport equation in flat-Earth and spherical-Earth geometries, and apply the method to Dirac dark matter with an isoscalar vector interaction. The iterative solution agrees well with the Monte Carlo spectra. \href{https://github.com/muling000/DM-earth-attenuation-boltzmann}{\faGithub}
}

\begin{document}
\maketitle
\flushbottom

\section{Introduction}
\label{sec:introduction}

Dark matter (DM) direct-detection experiments~\cite{Lin:2019uvt,Cooley:2021rws} are typically located deep underground to shield them from the overwhelming background of cosmic rays.
For DM with sufficiently small interaction cross sections, this shielding can usually
be ignored for DM signal, and the underground event rate may be computed from the incident
halo~\cite{McCabe:2010zh,Green:2011bv,Evans:2018bqy} or boosted-DM~\cite{Agashe:2014yua,Bringmann:2018cvk,Cappiello:2018hsu} flux.  This separation fails when the terrestrial overburden
is optically thick.  In searches for strongly interacting DM~\cite{Starkman:1990nj,Mahdawi:2017cxz,Davis:2017noy,Hooper:2018bfw,Emken:2018run}, light
DM~\cite{Emken:2017qmp,Cappiello:2023hza,Lantero-Barreda:2025sgy}, or boosted DM from cosmic rays and other energetic sources~\cite{Bringmann:2018cvk,Cappiello:2018hsu,Xia:2021vbz,Kolesova:2022kvq,Guha:2024mjr,Herbermann:2024kcy,DeRomeri:2023ytt,Jeesun:2026lro,Jeesun:2026wcx}, propagation through the Earth can
become part of the signal prediction itself, as emphasized by studies of terrestrial stopping~\cite{Starkman:1990nj,Kouvaris:2014lpa,Mahdawi:2017cxz,Davis:2017noy,Emken:2018run},
daily modulation~\cite{Collar:1992qc,Kouvaris:2014lpa,Ge:2020yuf,Chen:2021ifo,Qiao:2023pbw,Bertou:2025adb}, and underground scattering~\cite{Kavanagh:2016pyr,Emken:2017qmp,Kavanagh:2017cru,Eby:2023wem,CDEX:2021cll}.

The physical issue is more general than the loss of particles along a fixed
line of sight~\cite{Cappiello:2018hsu,Xia:2021vbz}.  Elastic scattering in the Earth removes particles from the
incident beam, but it also degrades their energy, changes their arrival
directions, and can scatter particles back into detector-accessible phase
space.  Thus, the same microscopic process produces both attenuation and
regeneration.  For an underground experiment, the relevant quantity is not only
a survival probability or a total attenuation factor.  One must determine how
the phase-space density evolves from the Earth's surface to the detector,
\begin{equation}
    f(\mathbf{x}_{\rm surf},\mathbf p_{\rm surf})
    \longrightarrow
    f(\mathbf{x}_{\rm det},\mathbf p_{\rm det}).
    \label{eq:intro_mapping}
\end{equation}
Here the left-hand side is evaluated for the incoming population at the surface,
and the right-hand side is the detector-side phase-space density that enters the
detector response.

This phase-space map is difficult to compute in full, so many treatments reduce
the propagation to a lower-dimensional estimate.  Optical-depth estimates
replace the passage through the Earth by a line integral of the inverse mean
free path, giving a survival factor for the unscattered beam~\cite{Bringmann:2018cvk,Cappiello:2018hsu,Xia:2021vbz,Kolesova:2022kvq,Guha:2024mjr,Herbermann:2024kcy}.
Alternatively, stopping-power approaches model the average energy loss along the trajectory to assess if
 the particle retains sufficient energy to produce a detectable signal~\cite{Kouvaris:2014lpa,Mahdawi:2017cxz,Emken:2018run}.
Such estimates give a useful first indication of shielding and upper reach, but
their output is essentially a path-dependent attenuation or energy-loss measure.
They do not by themselves reconstruct particles scattered into the detector
phase space.

More refined analytic treatments include part or all of the scattering
redistribution.  Single-scattering calculations integrate over the position and
kinematics of the first scattering, keeping both scatter-out and scatter-in as a
correction to pure survival \cite{Kavanagh:2016pyr}.  They apply when multiple
scatterings can be neglected.  Other semi-analytic approaches keep straight
chord trajectories but replace the full three-dimensional scattering history by
a forward/backward propagation map, treating attenuation and reflection with a
small number of scattering probabilities in specific light-DM models
\cite{Lantero-Barreda:2025sgy}.  Analytic multiple-scattering methods instead
avoid such reflected-trajectory bookkeeping by summing scattering orders
directly in light-DM limits \cite{Cappiello:2023hza}.  The efficiency of these
methods comes from reducing the full phase-space transport problem to a
lower-dimensional or model-specific calculation, but the same reductions also
limit their range of validity.

A complementary strategy employs trajectory-based Monte Carlo simulations.
By stochastically sampling the free path, target nucleus, and scattering kinematics, these methods track individual particles until they either penetrate the detector or are fully attenuated~\cite{Emken:2017qmp,Emken:2018run,Kavanagh:2017cru,DarkProp:v0.3}.
They are general and can handle complicated geometries and cross sections, but Monte Carlo methods are inherently limited by statistical noise in sparsely sampled regions of phase space.
Additionally, the sequential nature of event-by-event tracking renders them computationally inefficient when the optical depth is large.

In this work, we propose a Boltzmann transport formulation for Earth
attenuation.  The Boltzmann equation treats scatter-out and scatter-in through
the same collision operator.  Its integral solution naturally organizes the
propagated distribution as a sum over scattering orders~\cite{Ge:2024cto}, so multiple scattering
is obtained by accumulating successive scattering-order contributions rather
than by sampling individual trajectories.

The construction begins with the stationary-nuclei collision kernel, which maps
an incoming DM state into an outgoing energy and direction.  This keeps the
particle-physics input in the differential cross section and separates it from
the macroscopic propagation through the Earth.  We first discuss the flat-Earth
approximation, in which the relevant overburden is treated as a plane-parallel
slab.
This approximation is valid provided that the detector depth and transport length scales are small compared
with the Earth's radius.  In the isotropic light-DM limit, the spatial propagation and the
energy redistribution factorize, and the energy recurrence reduces to the
analytic propagation formula of Ref.~\cite{Cappiello:2023hza}.  We then derive
the spherical-Earth integral equation, whose straight-line characteristics are
chords through the Earth.  In the same factorizable limit, the chord geometry
sets the angular propagation while the energy spectrum follows the same
scattering-order recurrence as in the flat-Earth case.

For more general scattering kernels, the factorization need not hold.  If the
scattering is anisotropic in the Earth frame or if the outgoing energy and angle
are closely correlated by kinematics, the full distribution must be iterated in
energy and arrival direction simultaneously.  The Boltzmann integral equation
still gives such a scattering-order iteration.  As a benchmark, we apply the
formalism to Dirac DM with an isoscalar vector contact interaction.  The
resulting detector spectrum agrees very well with
the \textsc{DarkProp} Monte Carlo calculation~\cite{DarkProp:v0.3}
and the \textsc{DaMaSCUS} Monte Carlo calculation
~\cite{Emken:2018run,Emken:2017qmp}.

We organize the paper as follows.  In section~\ref{sec:setup} we present the
general Boltzmann equation and derive the stationary-nuclei collision kernel.  In
section~\ref{sec:flat} we discuss the flat-Earth approximation and its
factorized light-DM limit.
In section~\ref{sec:spherical} we derive the spherical-Earth transport
equation, its integral form, and the scattering-order solution.  In
section~\ref{sec:benchmark} we apply the method to a Dirac DM benchmark with an
isoscalar vector interaction.  Section~\ref{sec:discussion} concludes the
paper.  Numerical implementation details of the iteration are collected in
appendix~\ref{app:numerics}.

\section{Boltzmann equation and collision kernel}
\label{sec:setup}

Determining the DM flux relevant for an underground detector requires modeling the evolution of the incident surface distribution as it propagates through the Earth.
During this propagation, DM particles can elastically scatter from nuclei, resulting in energy loss and angular deflection.
The quantity to be transported therefore cannot be only a total flux
or a direction-independent attenuation factor. It must retain the particle
position and momentum.  We describe this
state by the phase-space distribution $f(\mathbf{x},\mathbf p,t)$ of the DM
particle $\chi$, normalized to the number of particles $N$ in the corresponding
phase-space element as
\begin{equation}
    dN
    =
    g_\chi
    f(\mathbf{x},\mathbf{p},t)
    \frac{d^3\mathbf{x}\,d^3\mathbf{p}}{(2\pi)^3}.
\end{equation}
Here $g_\chi$ denotes the internal degrees of freedom of the DM particle.

The propagation of this phase-space distribution is governed by a Boltzmann
equation, which can be written schematically as
\begin{equation}
    \hat{\mathbf{L}}[f]
    =
    \mathbf{C}[f].
\end{equation}
Here $\hat{\mathbf{L}}[f]$ is the Liouville operator and
$\mathbf{C}[f]$ is the collision operator.
For the flat-spacetime propagation considered here, the Liouville operator, or
free-streaming part, is~\cite{Kolb:1990vq}
\begin{equation}
    \hat{\boldsymbol{\mathrm{L}}}
    =
    E\frac{\partial}{\partial t}
    +
    \mathbf p\cdot\nabla_{\mathbf x}.
    \label{eq:flat_liouville}
\end{equation}
The transport equations below use the steady-state limit, in which $f$ has no
explicit time dependence and the $\partial/\partial t$ term in
Eq.~\eqref{eq:flat_liouville} is dropped.  The Liouville term is specialized
separately to the flat-Earth and spherical-Earth geometries in the following.

The other side of the Boltzmann equation encodes collisions.  The invariant collision operator for a general
process $\chi+a\leftrightarrow i+j$ is~\cite{Kolb:1990vq}
\begin{equation}
\begin{aligned}
    \boldsymbol{\mathrm{C}}[f_\chi]
    &=
    -\frac{1}{2}
    \int d\Pi_a\,d\Pi_i\,d\Pi_j\,
    (2\pi)^4\delta^4(p+p_a-p_i-p_j)
    \\
    &\quad\times
    \left[
    \overline{\overline{|\mathcal M|_{\chi a\to ij}^2}}\,
    f_\chi f_a(1\pm f_i)(1\pm f_j)
    -
    \overline{\overline{|\mathcal M|_{ij\to\chi a}^2}}\,
    f_i f_j(1\pm f_\chi)(1\pm f_a)
    \right],
\end{aligned}
    \label{eq:collision_general_f}
\end{equation}
where the invariant phase-space measure is
\begin{equation}
    d\Pi_s
    =
    \frac{g_s}{(2\pi)^3}\frac{d^3\mathbf p_s}{2E_s},
\end{equation}
$g_s$ is the internal degeneracy of
species $s$, $\mathcal M$ is the invariant scattering amplitude, and the
upper/lower sign applies to bosons/fermions.  The double
overline indicates that the squared amplitude is averaged over initial and
final spin states and includes the appropriate identical-particle symmetry
factors. 

For elastic scattering on a target species $A$,
$\chi + A\to\chi' + A'$, assuming $T$ (or $CP$) invariance so that the squared
amplitudes for the forward and inverse processes are equal, and neglecting Bose enhancement
and Pauli blocking by setting $1\pm f\simeq1$, the collision operator reduces to
the following form:
\begin{equation}
    \begin{aligned}
        \boldsymbol{\mathrm{C}}_{\chi A}[f_\chi]
        &=
        -\frac{1}{2}
        \int d\Pi_A d\Pi_{\chi'} d\Pi_{A'}
        (2\pi)^4 \delta^4(p+p_A-p'-p'_A)
        \overline{\overline{|\mathcal M|_{\chi A}^2}}
        \left(f_\chi f_A-f_{\chi'}f_{A'}\right)
        \\
        &\equiv
        \boldsymbol{\mathrm{C}}_{\chi A}^{\mathrm{loss}}[f_\chi]
        +
        \boldsymbol{\mathrm{C}}_{\chi A}^{\mathrm{gain}}[f_\chi].
    \end{aligned}
    \label{eq:collision_elastic_split}
\end{equation}
Here $f_\chi$ and $f_A$ refer to the phase-space densities at the initial
four-momenta $p$ and $p_A$, while $f_{\chi'}$ and $f_{A'}$ refer to those at the
final four-momenta $p'$ and $p'_A$.
The loss term $\boldsymbol{\mathrm{C}}_{\chi A}^{\mathrm{loss}}[f_\chi]$ removes particles from the observed phase-space state by
scattering them into other energies or directions.  The gain term $\boldsymbol{\mathrm{C}}_{\chi A}^{\mathrm{gain}}[f_\chi]$ repopulates
the same state by scattering particles from other incoming states into
$(\mathbf x,\mathbf p)$.

Inside the Earth, the thermal motion of the target nuclei is negligible on the
DM scattering scales of interest, so we approximate them as stationary
scattering centers.  For a target species $A$ with number density $n_A$,
\begin{equation}
    f_A
    =
    n_A\frac{(2\pi)^3}{g_A}\delta^3(\mathbf p_A).
\end{equation}
The loss term is then fixed by the total cross section for a DM particle in
the observed state,
\begin{equation}
    \boldsymbol{\mathrm{C}}_{\chi A}^{\mathrm{loss}}[f_\chi]
    =
    -|\mathbf p|\,f_\chi\,n_A\sigma_A .
    \label{eq:collision_loss_stationary}
\end{equation}
The gain term is most conveniently written in terms of the Earth-frame
differential cross section for an incoming state $(E',\Omega')$ to scatter into
the observed state $(E,\Omega)$,
\begin{equation}
    \boldsymbol{\mathrm{C}}_{\chi A}^{\mathrm{gain}}[f_\chi]
    =
    \frac{n_A}{|\mathbf p|}
    \int dE' d\Omega'\,
    \frac{d\sigma_A}{dE\,d\Omega}
    |\mathbf p'|^2 f_\chi(\mathbf x,\mathbf p').
    \label{eq:collision_gain_stationary}
\end{equation}
Here $d\sigma_A/dE\,d\Omega$ denotes the differential cross section in the
Earth frame.  For elastic scattering from a stationary target, two-body
kinematics restricts the allowed outgoing energy for each incoming energy
$E'$. Thus, at fixed $E$, the scatter-in integral receives contributions only
from incoming energies $E'$ that can scatter to $E$.

For several nuclear species, the two collision terms are summed over all nuclear species $A$ in the Earth.
With a constant mean free path,
$
    l^{-1}
    \equiv
    \sum_A n_A\sigma_A ,
$
the collision operator contributes
\begin{subequations}
\label{eq:stationary_collision_terms}
    \begin{align}
        \sum_A \boldsymbol{\mathrm{C}}_{\chi A}^{\mathrm{loss}}[f_\chi]
        &=
        -\frac{1}{l}|\mathbf p| f_\chi ,     \\
        \sum_A \boldsymbol{\mathrm{C}}_{\chi A}^{\mathrm{gain}}[f_\chi]
        &=
        \frac{1}{|\mathbf p|}
        \sum_A n_A
        \int dE' d\Omega'\,
        \frac{d\sigma_A}{dE\,d\Omega}
        |\mathbf p'|^2 f_\chi(\mathbf x,\mathbf p') .
    \end{align} %
\end{subequations} %
If the total cross section is energy dependent, the same structure applies with
$l$ replaced by the appropriate energy-dependent attenuation length.

\section{The flat-Earth approximation}
\label{sec:flat}

When the detector depth and the mean free path $l$ are both much smaller than
the Earth's radius, Earth curvature can be neglected, and the medium can be
approximated as an infinite plate.  If $l$ is also shorter than the detector
depth, the detector is primarily sensitive to particles repopulated by
scatter-in from the nearby overburden.
We take $z=0$ at the surface, $z>0$ into the Earth, and define the direction cosine
$u \equiv \cos\theta$, where $\theta$ is the polar angle of the particle's momentum relative to the downward vertical. Translational
symmetry in the transverse directions and axial symmetry around the $z$ axis
reduce the distribution to $f_\chi (\mathbf{x}, \mathbf{p}) = f_\chi(z,u,E)$. By absorbing the phase-space factor $|\mathbf p|^2$ from the integration measure $d^3\mathbf p = |\mathbf p|^2 dp d\Omega$ into the definition
\begin{equation}
    F(z,u,E)
    \equiv
    |\mathbf p|^2 f_\chi(z,u,E),
\end{equation}
the Boltzmann transport equation simplifies to
\begin{equation}
    u\,\frac{\partial F(z,u,E)}{\partial z}
    +
    \frac{1}{l}F(z,u,E)
    =
    \sum_A n_A
    \int dE' d\Omega'
    \frac{d\sigma_A}{dE\,d\Omega}
    F(z,u',E').
    \label{eq:plate_transport}
\end{equation}
The first term streams the distribution along a straight ray, the second term
is scatter-out with mean free path $l$, and the right-hand side is the local
scatter-in source from other directions and energies.
We denote the prescribed incident DM distribution at the surface for
downward-moving particles by $G(0,u,E)$.  The boundary conditions are
\begin{subequations}
\label{eq:plate_boundary}
\begin{alignat}{2}
    F(0,u,E) &= G(0,u,E) && \quad \text{for} \quad 0 < u \le 1, \\
    F(\infty,u,E) &= 0 && \quad \text{for} \quad -1 \le u \le 0,
\end{alignat}
\end{subequations}
The first condition matches $F$ to this surface distribution, while the second
imposes no incoming beam from infinite depth, so upward
moving particles are generated only by scatter-in.

For a constant mean free path $l$, Eq.~\eqref{eq:plate_transport} can be integrated directly along the straight-line trajectories of the particles, yielding an integral equation for $F$:
\begin{subequations}
\label{eq:plate_integral_branches}
    \begin{align}
        F(z,u>0,E)
        &=
        G(0,u,E) e^{-z/(lu)}
        \notag\\
        &\quad
        +
        \frac{1}{|u|}
        \int_0^z dz'\,
        e^{-|z-z'|/(l|u|)}
        \sum_A n_A
        \int dE' d\Omega'\,
        \frac{d\sigma_A}{dE\,d\Omega}
        F(z',u',E'),
        \\
        F(z,u<0,E)
        &=
        \frac{1}{|u|}
        \int_z^\infty dz'\,
        e^{-|z-z'|/(l|u|)} 
        \sum_A n_A
        \int dE' d\Omega'\,
        \frac{d\sigma_A}{dE\,d\Omega}
        F(z',u',E').
    \end{align}
\end{subequations}
The first line contains the unscattered surface contribution plus all particles
whose last scattering occurred between the surface and the observation point.
The second line has no boundary term because the upward-moving component is
entirely generated inside the plate.

For light DM ($m_\chi \ll m_A$), assuming the scattering is isotropic in the center-of-mass (CoM) frame, the CoM frame approximately coincides with the Earth frame, and the maximum energy transfer fraction is small. Consequently, the scattering remains nearly isotropic in the Earth frame, and the final energy of the scattered particle becomes nearly decoupled from its scattering angle~\cite{Cappiello:2023hza}. Under these conditions, the differential cross section can be approximated as
\begin{equation}
    \frac{d\sigma_A}{dE\,d\Omega}
    \simeq
    \frac{1}{4\pi}\frac{d\sigma_A}{dE}
    \simeq
    \frac{1}{4\pi}\frac{\sigma_A}{\Delta E_{\max}},
    \label{eq:plate_isotropic_kernel}
\end{equation}
where $\Delta E_{\max}$ is the maximum kinematic energy transfer.
In this case, we can integrate out $u$ in Eq.~\eqref{eq:plate_integral_branches}
to obtain the integral equation for the differential flux, which is related to
$F$ by $d\Phi/dE=(2\pi)^{-3}\int d\Omega\,F(z,u,E)$.
Assuming the incoming dark matter is also isotropic, $G(0,u,E)=G(0,E)$, the
energy-dependent flux satisfies the integral equation
\begin{equation}
    \frac{d\Phi}{dE} (z,E) 
    =
    \frac{E_2(z/l) }{4\pi^2} G(0,E) 
    + 
    \frac{1}{2}
    \int_0^\infty dz'
    \Gamma(0, |z-z'|/l)
    \sum_A 
    \int dE' 
    n_A \frac{d\sigma_A}{dE}
    \frac{d\Phi}{dE'} (z', E') .
    \label{eq:plate_energy_flux_integral}
\end{equation}
where $E_2(x)$ is the generalized exponential integral and
$\Gamma(0,x)$ is the upper incomplete gamma function.
This equation can be solved by expanding the differential flux in a Neumann series, $\frac{d\Phi}{dE} = \sum_{i=0}^{\infty} \frac{d\Phi_i}{dE}$, where the differential flux of the $i$-th scattering order satisfies the recurrence relation
\begin{subequations}
    \label{eq:plate_differential_flux_recursion}
\begin{align}
    \frac{d\Phi_0}{dE} (z,E)
    &=
    \frac{1}{4\pi^2} G(0,E) E_2(z/l), \\
    \frac{d\Phi_{i+1}}{dE} (z,E)
    &=
    \frac{1}{2} \int_0^\infty dz' \Gamma \left(0, |z-z'|/l \right) \sum_A \int dE' n_A \frac{d\sigma_A}{dE} \frac{d\Phi_i}{dE'} (z',E').
\end{align}
\end{subequations}
Here, $\frac{d\Phi_0}{dE}(z, E)$ represents the unscattered differential flux. For the $(i+1)$-th order contribution, the flux at depth $z$ is built up by the scattering of the $i$-th order flux at all other depths $z'$, and the spatial propagation after scattering is described by the kernel $\Gamma(0, |z-z'|/l)$.

Because the spatial propagation and energy scattering kernels are decoupled, the spatial and energy dependence of the differential flux in Eq.~\eqref{eq:plate_differential_flux_recursion} can be separated. Specifically, the energy-dependent flux of the $i$-th scattering order factorizes as
\begin{equation}
    \frac{d\Phi_i}{dE} (z,E)
    =
    \Phi_i(z) \frac{dN_i}{dE} (E),
    \label{eq:plate_flux_factorization}
\end{equation}
where $\Phi_i(z)$ is the total flux of the $i$-th scattering order, and $\frac{dN_i}{dE} (E)$ is the normalized energy spectrum.

\begin{figure}[t]
    \centering
    \includegraphics[width=0.62\textwidth]{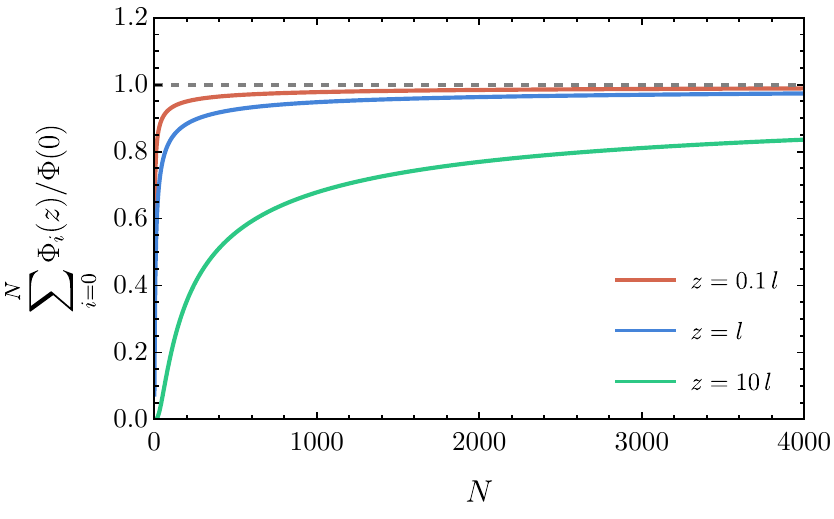}
    \caption{The normalized total flux of DM at depths $z=0.1\,l$ (red), $l$ (blue), and $10\,l$ (green) in the flat-Earth approximation. The sum of scattering-order contributions converges to the analytical constant total flux given in Eq.~\eqref{eq:plate_constant_flux} as the maximum scattering order $N$ increases.}
    \label{fig:iteration_flux_plate}
\end{figure}

The integral equation satisfied by the total flux $\Phi(z)$ is obtained by integrating Eq.~\eqref{eq:plate_energy_flux_integral} over all energies $E$:
\begin{equation}
    \Phi(z)
    =
    \frac{1}{2} \Phi(0) E_2(z/l)
    +
    \frac{1}{2}
    \int_0^\infty \frac{dz'}{l}
    \Gamma(0, |z-z'|/l)
    \Phi(z'),
    \label{eq:plate_flux_integral}
\end{equation}
This equation admits the constant solution
\begin{equation}
    \Phi(z) = \Phi(0)
    =
    \frac{1}{2\pi^2}\int dE\,G(0,E)
    \label{eq:plate_constant_flux}
\end{equation}
because the loss and gain terms balance after integration over angle and energy. This integral equation also admits a Neumann series solution $\Phi(z) = \sum_{i=0}^{\infty} \Phi_i (z)$, where the spatial component of each scattering order satisfies the iteration relation
\begin{subequations}
\label{eq:plate_spatial_flux_recursion}
\begin{align}
    \Phi_0(z)
    &=
    \frac{1}{2} \Phi(0) E_2(z/l), \\
    \Phi_{i+1}(z)
    &=
    \frac{1}{2} \int_0^\infty \frac{dz'}{l} \Gamma \left(0, |z-z'|/l \right) \Phi_i(z'),
\end{align}
\end{subequations}
This reproduces the recurrence relation of Ref.~\cite{Cappiello:2023hza}.
The scattering-order solution recovers the analytical constant total flux result in Eq.~\eqref{eq:plate_constant_flux}, as shown in Fig.~\ref{fig:iteration_flux_plate}. In the figure, we show the accumulated total flux at three different depths, $z = 0.1\,l$, $z=l$, and $z=10\,l$, as a function of the maximum scattering order $N$. When $N$ is large, this sum converges to the constant analytical value, with faster convergence at smaller $z$ where fewer scatterings are needed.

To obtain the detailed energy distribution, we solve the energy-dependent flux equation \eqref{eq:plate_energy_flux_integral} using the scattering-order method. Based on the factorization in Eq.~\eqref{eq:plate_flux_factorization}, the normalized energy spectrum satisfies the iteration relation
\begin{subequations}
\label{eq:plate_spectrum_recursion}
\begin{align}
    \frac{dN_0}{dE} (E)
    &=
    \frac{G(0,E)}{\int dE G(0,E)}, \\
    \frac{dN_{i+1}}{dE} (E)
    &=
    \sum_A \int dE' n_A l \frac{d\sigma_A}{dE} \frac{dN_i}{dE} (E').
\end{align}
\end{subequations}
One can verify that the normalization of $dN_i / dE$ is preserved at each step. 
The scattering-order approach describes how the energy-dependent flux builds up order-by-order, analogous to the successive approximation method in classical radiative transfer problems~\cite{Chandrasekhar:1960}.
In the non-relativistic limit, by introducing the variable substitution $\delta T \equiv (T'-T) T/T'$, where $T = E - m_\chi$ and $T' = E' - m_\chi$ are the kinetic energies, this recurrence relation for the spectrum reduces to the iterative propagation formula in Ref.~\cite{Cappiello:2023hza}. 

\begin{figure}[t]
    \centering
    \includegraphics[width=0.78\textwidth]{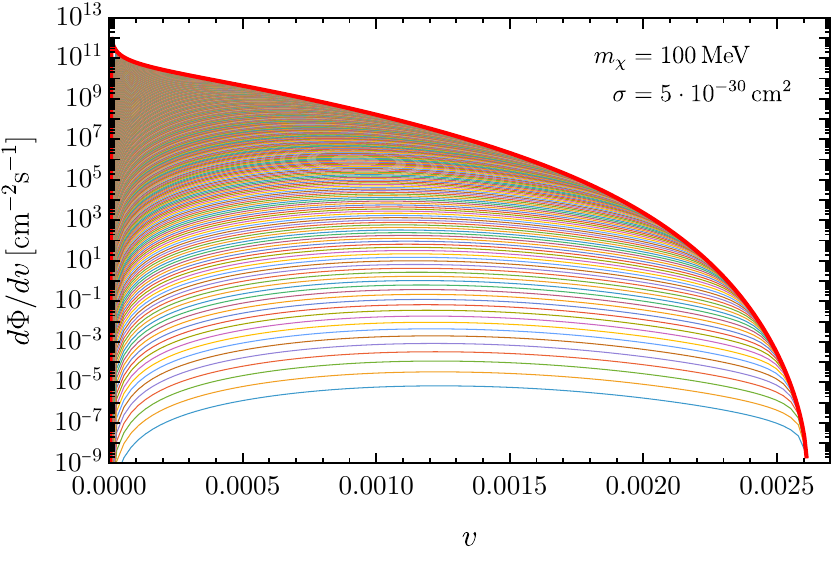}
    \caption{DM spectrum at the Jinping Underground Laboratory depth in the flat-Earth approximation. The propagated dark matter spectrum with $m_\chi = 100\text{ MeV}$ and $\sigma = 5\times 10^{-30}\text{ cm}^2$ is shown as a function of DM velocity $v$, demonstrating convergence as the number of scatterings increases.}
    \label{fig:plate_methods}
\end{figure}

Fig.~\ref{fig:plate_methods} shows the propagated spectrum obtained from the iterative scattering-order method. We choose a dark matter mass of $m_\chi = 100\text{ MeV}$, a DM-nucleon scattering cross section of $\sigma = 5\times 10^{-30}\text{ cm}^2$, and a depth of $z = 2.4\text{ km}$, corresponding to the depth of the China Jinping Underground Laboratory~\cite{PandaX:2014mem,CDEX:2021cll}. At this depth, the high-energy part of the spectrum is dominated by particles that have undergone few scatterings, resulting in a suppressed flux. In contrast, particles at lower energies have experienced a large number of collisions, reaching the order of $10^3$ scatterings. 

\section{The spherical-Earth geometry}
\label{sec:spherical}

In practice, an underground detector is not exclusively sensitive to dark matter
traversing the local vertical overburden.
Particles can enter the Earth from different
surface locations, travel over different chord lengths, and scatter inside the
Earth before reaching the detector.  The Earth therefore acts as a finite
scattering medium that redistributes the incoming population in both energy and
arrival direction.  The flat-Earth approximation in section~\ref{sec:flat}
captures the local limit of this process, while the full detector distribution
requires the distribution-function map to be formulated in
spherical geometry.

To set up this map, we use a spherically symmetric, time-independent Earth
model.  The phase-space distribution is then independent of
the spatial angles and of the azimuthal momentum angle.  It is specified by the radius $r$, the local direction cosine
$u$, and the energy $E$.  As in the preceding sections, we absorb the
factor $|\mathbf p|^2$ from the momentum measure into
$F(r,u,E) \equiv |\mathbf p|^2 f_\chi(r,u,E)$.

In these variables, free streaming through the spherical medium is described by
the Liouville operator
\begin{equation}
  \hat{\boldsymbol{\mathrm{L}}}[F]
  =
  u\,\frac{\partial F}{\partial r}
  +
  \frac{1-u^2}{r}\frac{\partial F}{\partial u}.
  \label{eq:spherical_liouville}
\end{equation}
The second term is purely geometric: even for straight-line
propagation, the local angle relative to the radial direction changes along a
chord.  With stationary target nuclei and a constant mean free path
$l^{-1}\equiv\sum_A n_A\sigma_A$, the spherical transport equation is
\begin{equation}
  u \frac{\partial F(r,u,E)}{\partial r} 
  + 
  \frac{1-u^2}{r} \frac{\partial F(r,u,E)}{\partial u} 
  + 
  \frac{1}{l} 
  F(r,u,E) 
  =
  \sum_A n_A \int dE' d\Omega' \, \frac{d\sigma_A}{dE\,d\Omega} F(r,u',E'),
  \label{eq:spherical_transport_eq}
\end{equation}
where $u'$ and $E'$ are the direction cosine and energy before the scattering
that produces a particle with energy $E$ and direction cosine $u$. 

The characteristic curves are straight chords.  Along each chord, it is useful
to replace $(r,u)$ by $x \equiv r u$ and
$y \equiv r \sqrt{1-u^2}$, where $y$ is the impact parameter and $x$ is the
coordinate along the chord.  The radial distance at any point on the chord is
$r(x,y) = \sqrt{x^2+y^2}$.
At fixed $y$, the streaming operator becomes a derivative with respect to $x$.
Tracing the chord backward to the surface gives
$x=-x_\oplus(y)$, with
$x_\oplus(y)\equiv\sqrt{R_\oplus^2-y^2}$, and the corresponding surface
direction cosine is $u_\oplus=-x_\oplus(y)/R_\oplus$.  The required
boundary data are the incoming surface distribution at this upstream point,
denoted by $G(R_\oplus,u_\oplus,E)$.  Integrating
Eq.~\eqref{eq:spherical_transport_eq} from this boundary point to the
observation point gives
\begin{equation}
\begin{aligned}
    F(r,u,E)
    ={}&
    G(R_\oplus, u_\oplus, E) e^{-(x+x_\oplus)/l}
    \\
    &+
    \int_{-x_\oplus}^x \frac{dx'}{l} \,
    e^{-(x-x')/l}
    \sum_A n_A l
    \int dE' d\Omega' \,
    \frac{d\sigma_A}{dE\,d\Omega}
    F(r(x',y), u', E') .
\end{aligned}
    \label{eq:spherical_integral}
\end{equation}
The second line in Eq.~\eqref{eq:spherical_integral} sums over the possible
last-scattering point $x'$.  The symbol $u'$ denotes the incoming direction
before that scattering.  The outgoing direction at that point is not the
detector direction cosine $u=x/r$, but the local chord direction, which reduces
to $u$ at the endpoint $x'=x$.  This distinction is essential in spherical geometry
because the local radial vector changes along the chord.  Thus,
Eq.~\eqref{eq:spherical_integral} is the explicit spherical realization of the
transport map from the surface distribution to that at the detector depth.

\begin{figure}[t]
    \centering
    \includegraphics[width=0.62\textwidth]{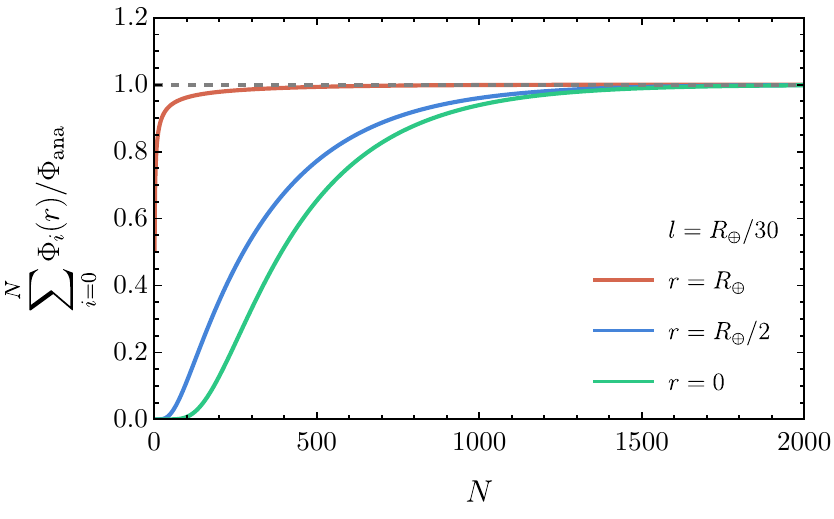}
    \caption{The normalized total flux of dark matter at the surface (red), at half the Earth
    radius (blue), and at the Earth center (green) in the spherical-Earth approximation.  The numerical Neumann solution converges to the exact constant analytic result as the maximum scattering order $N$ increases.}
    \label{fig:spherical_flux_validation}
\end{figure}

For light DM with isotropic scattering in the center-of-mass frame, we use the
same approximation of differential cross section as in Eq.~\eqref{eq:plate_isotropic_kernel}.  Assuming an isotropic incident
flux, the surface distribution reduces to $G(R_\oplus,E)$.  In this case, the
variables $(r,u)$ and $E$ factorize, and the integral equation for the
differential flux in $(r,u)$ can be obtained by integrating
Eq.~\eqref{eq:spherical_integral} over energy.  This yields
\begin{equation}
    \frac{d\Phi}{du}(r,u) 
    =
    \frac{e^{-(x + x_\oplus) / l}}{4\pi^2} \int dE \, G(R_\oplus,E) 
    +
    \int_{-x_\oplus}^x \frac{dx'}{l} \,
    e^{-(x-x')/l} 
    \int \frac{du'}{2}
    \frac{d\Phi}{du} (r(x',y),u'),
    \label{eq:spherical_flux_integral_eq}
\end{equation}
This integral equation admits a constant analytic solution for the angular flux,
\begin{equation}
    \frac{d\Phi}{du}(r,u) = \frac{1}{4\pi^2} \int dE \, G(R_\oplus,E).
    \label{eq:spherical_constant_flux}
\end{equation}
Substituting this constant solution into the scattering term effectively compensates for the missing part of the attenuated boundary term.
The corresponding scattering-order recurrence solution of the angular flux is
\begin{subequations}
\label{eq:spherical_flux_recursion}
\begin{align}
    \frac{d\Phi_0}{du}(r,u)
    &=
    \frac{e^{-(x + x_\oplus) / l}}{4\pi^2} \int dE \, G(R_\oplus,E) ,
    \\
    \frac{d\Phi_{i+1}}{du} (r,u)
    &=
    \int_{-x_\oplus}^x \frac{dx'}{l} \,
    e^{-(x-x')/l}
    \int \frac{du'}{2}
    \frac{d\Phi_i}{du} (r(x',y),u').
\end{align}
\end{subequations}
The numerical scattering-order solution is shown in Fig.~\ref{fig:spherical_flux_validation}.  In this figure, we plot the total flux at radial distance $r$, obtained by integrating the angular flux over the direction cosine.  We show the flux evaluated at the Earth center, at half the Earth radius, and at the surface.  When the total number of scatterings $N$ is large, the numerical summation converges to the constant analytic result $\Phi_\mathrm{ana}$.

\begin{figure}[t]
    \centering
    \includegraphics[width=0.78\textwidth]{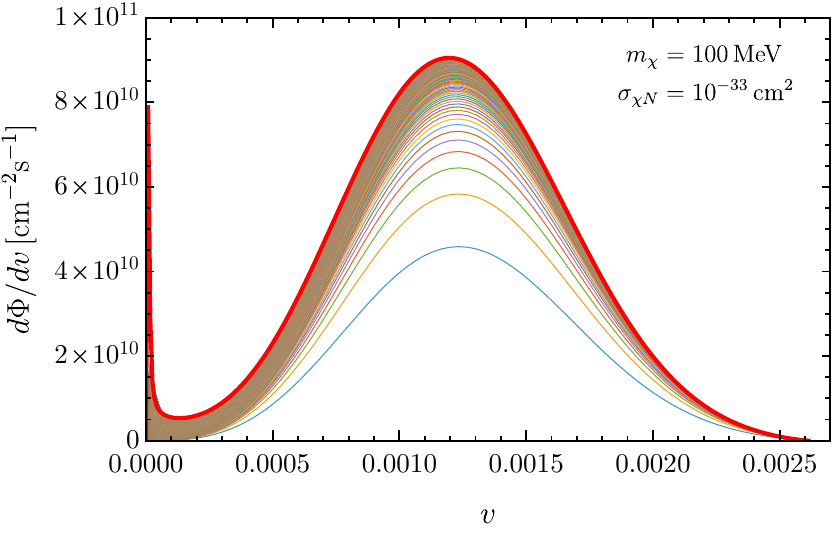}
    \caption{The dark matter spectrum at the Jinping Underground
    Laboratory depth in the spherical-Earth approximation with $m_\chi = 100\,\text{MeV}$ and $\sigma = 10^{-33}\,\text{cm}^2$.}
    \label{fig:spherical_spectrum}
\end{figure}

While the numerical summation confirms that the spatial distribution converges to
a constant flux, the energy spectrum continues to evolve with each collision.
In the isotropic limit, the collision kernel decouples the scattering angle from
the energy transfer, and the phase space density factorizes, similar to Eq.~\eqref{eq:plate_flux_factorization}.  Under
this factorization, the normalized energy spectrum $dN_i/dE$ can be computed
independently via the recurrence relation
\begin{subequations}
\label{eq:spherical_spectrum_recursion}
\begin{align}
    \frac{dN_0}{dE}(E) &= \frac{G(R_\oplus,E)}{\int dE' \, G(R_\oplus,E')}, \\
    \frac{dN_{i+1}}{dE}(E) &= \sum_A \int dE' \, n_A l \frac{d\sigma_A}{dE} \frac{dN_i}{dE}(E').
\end{align}
\end{subequations}
The energy recurrence has the same structure as Eq.~\eqref{eq:plate_spectrum_recursion} in the flat-Earth approximation, while
the angular weight is now determined by the spherical chord geometry.  The
propagated spectrum obtained by summing the scattering-order contributions is
shown in Fig.~\ref{fig:spherical_spectrum}.

\section{Vector-coupling model and non-factorizable scattering}
\label{sec:benchmark}

The spherical transport equation derived above separates the macroscopic
propagation problem from the microscopic scattering input.  The chord geometry
determines how particles move through the Earth, while the particle model enters
through the differential scattering cross section.  The isotropic limits used above assume that
scattering is approximately isotropic in the Earth frame and that the final
energy of the scattered particle is decoupled from its scattering angle.  This
can be a useful approximation for light DM, $m_\chi\ll m_A$, when an isotropic
center-of-mass distribution remains nearly isotropic in the Earth frame and
the maximum fractional energy transfer is small.

A more general model requires a different treatment for two reasons.
First, the scattering kernel need not be isotropic in the Earth frame.
Second, for heavier DM the energy loss can be large enough that the final
energy of the scattered particle can no longer be approximated as independent
of its outgoing direction.  In this case, the full differential kernel must be
kept. 

We illustrate this general case with a Dirac fermion $\chi$ with an isoscalar
vector contact interaction~\cite{Fitzpatrick:2012ix,Anand:2013yka},
\begin{equation}
    \mathcal L_{\rm eff}
    =
    \frac{1}{\Lambda^2}
    (\bar\chi\gamma^\mu\chi)
    (\bar p\gamma_\mu p+\bar n\gamma_\mu n).
    \label{eq:benchmark_operator}
\end{equation}
For a nucleus of mass $m_A$ and mass number $\mathcal A$, neglecting nuclear
form factors, the spin-averaged squared matrix element in the Earth frame is
\begin{equation}
    \overline{|\mathcal M_{\chi A}|^2}
    =
    \frac{8\mathcal A^2}{\Lambda^4}
    \left[
    m_A^2(E'^2+E^2)
    -
    m_A(E'-E)(m_\chi^2+m_A^2)
    \right].
    \label{eq:matrix_element_benchmark_vector_earth}
\end{equation}
Here $E'$ and $E$ are the incoming and outgoing DM energies, respectively.
The differential cross section entering the
Boltzmann equation is
\begin{equation}
    \frac{d\sigma_A}{dE\,d\Omega}
    =
    \frac{1}{64\pi^2m_A|\mathbf p'|^2}
    \overline{|\mathcal M_{\chi A}|^2}
    \delta\!\left(
    \cos\theta_{\chi\chi'}
    -
    C_A(E,E')
    \right),
    \label{eq:benchmark_dsigma}
\end{equation}
where $\theta_{\chi\chi'}$ is the scattering angle of the dark matter particle
in the Earth frame.  The kinematic correlation factor is
\begin{equation}
    C_A(E,E')
    =
    \frac{EE'-m_\chi^2+m_A(E-E')}
    {|\mathbf p|\,|\mathbf p'|}.
    \label{eq:benchmark_C}
\end{equation}
The $\delta$-function in Eq.~\eqref{eq:benchmark_dsigma} fixes the scattering angle for given incoming and outgoing energies.  This is the kinematic correlation that prevents the general Earth-frame kernel from factorizing into independent energy and angular pieces.

\begin{figure}[t]
    \centering
    \includegraphics[width=0.78\textwidth]{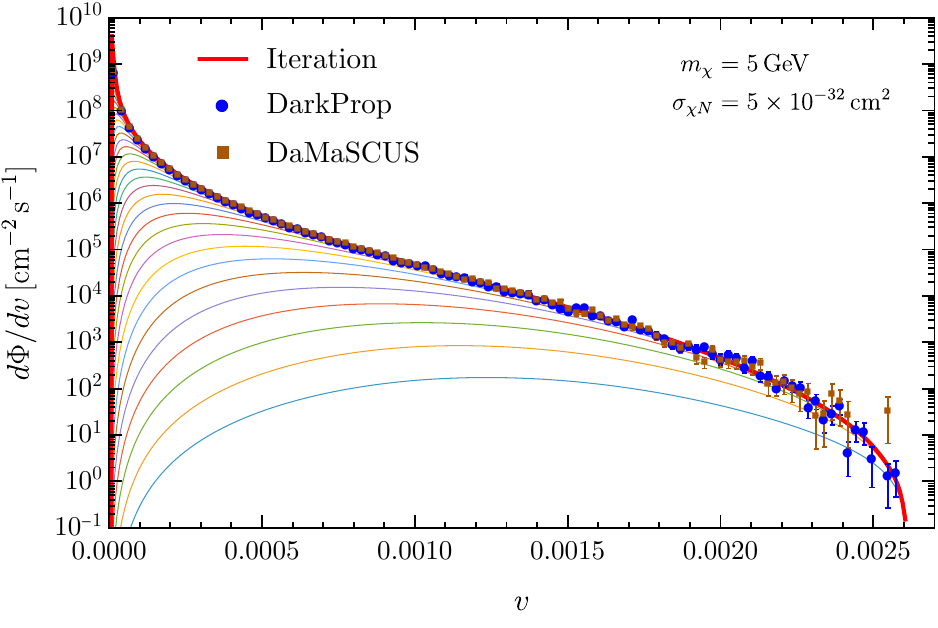}
    \caption{Propagated detector spectrum for the Dirac DM benchmark with an
    isoscalar vector coupling, evaluated at
    $m_\chi=5\,\mathrm{GeV}$,
    $\sigma_{\chi N}=5\times10^{-32}\,\mathrm{cm}^2$, detector depth
    $2.4\,\mathrm{km}$, and an isotropic truncated-Maxwellian surface spectrum~\cite{Baxter:2021pqo}.  The
    red curve is the scattering-order iteration result, and the blue and brown
    points show the corresponding \textsc{DarkProp} and \textsc{DaMaSCUS}
    Monte Carlo results, respectively.}
    \label{fig:dirac_benchmark_spectrum}
\end{figure}

Since the variables $(r,u)$ and $E$ no longer separate, we solve the full
spherical integral equation \eqref{eq:spherical_integral} as an iteration for the distribution
$F(r,u,E)$ itself, with $F(r,u,E)=\sum_i F_i(r,u,E)$.  For isotropic
incident boundary conditions, the surface distribution is
$G(R_\oplus,E)$, and the scattering orders are given by
\begin{subequations}
\label{eq:neumann_series_spherical}
\begin{align}
    F_0(r,u,E)
    &=
    G(R_\oplus,E) e^{-(x+x_\oplus)/l},
    \\
    F_{i+1}(r,u,E)
    &=
    \int_{-x_\oplus}^{x}\frac{dx'}{l}\,
    e^{-(x-x')/l}
    \sum_A n_A l
    \int dE' d\Omega'\,
    \frac{d\sigma_A}{dE\,d\Omega}
    F_i(r(x',y),u',E').
\end{align}
\end{subequations}
Fig.~\ref{fig:dirac_benchmark_spectrum} shows the solution of this iteration
for $m_\chi=5\,\mathrm{GeV}$ and
$\sigma_{\chi N}=5\times10^{-32}\,\mathrm{cm}^2$.  Because the mass is much
larger than in the light-DM example, each scattering removes a
larger fraction of the incoming kinetic energy.  The propagated spectrum
therefore converges after noticeably fewer scattering orders.

The figure also compares the iterative result with
the Monte Carlo spectra obtained with
\textsc{DarkProp}~\cite{DarkProp:v0.3} and
\textsc{DaMaSCUS}~\cite{Emken:2018run,Emken:2017qmp}.  The
three calculations agree very well.
At the high-velocity end of the propagated spectrum,
the Monte Carlo samples contain fewer events and therefore have larger
statistical uncertainties, but remain consistent with the iterative result.  The
iteration method does not rely on event sampling, and it gives a
more accurate spectrum there.
A separate advantage appears at large cross sections.  A Monte Carlo simulation
then becomes inefficient because few particles survive to the detector.  The
iteration method is controlled mainly by the number of scattering orders needed
for energy degradation.  This number is set by the energy loss per collision
and does not grow with the total cross section, so the calculation remains fast
even at large cross section.

We use Dirac DM with an isoscalar vector interaction to benchmark non-factorizable transport.
The formalism can be applied to other elastic DM--nucleus interactions by
replacing the corresponding differential scattering kernel and
interaction-dependent mean free path in Eq.~\eqref{eq:neumann_series_spherical}.

\section{Conclusion}
\label{sec:discussion}

In this work, we have formulated dark matter propagation through the Earth as a
Boltzmann transport problem for the detector phase-space distribution.  The
formulation treats scatter-out and scatter-in in the same collision operator, so
attenuation, energy degradation, angular redistribution, and regeneration are
described by a surface-to-detector map rather than by a survival probability
alone.  In the flat-Earth and spherical-Earth limits, the light-DM isotropic case
provides analytic checks: the total flux has a constant analytic solution, while
the energy spectrum evolves order by order through the same recurrence
structure.  For non-factorizable kernels, illustrated by a Dirac dark matter
benchmark with an isoscalar vector contact interaction, the full distribution
function can be iterated directly and gives very good agreement with
\textsc{DarkProp} and \textsc{DaMaSCUS}.  The same framework can be systematically generalized beyond the simplifying
assumptions used here --- stationary targets, spherical symmetry, and a
constant mean free path --- by incorporating realistic density profiles,
energy-dependent attenuation lengths, and nuclear form factors.  By applying this method to strongly interacting dark matter, we can better translate direct-detection data into reliable constraints on dark matter interactions.

\appendix

\section{Numerical implementation}
\label{app:numerics}

The scattering-order recurrences in the main text share a similar form.  Each order is obtained by applying an integral operator to the
previous one, so the numerical task is to discretize that operator.  Quadrature
evaluates the integral at sampled points, and interpolation maps the sampled
values back to the chosen grid.  After this discretization, advancing by one
scattering order is a matrix multiplication.

Suppose an integral operator is evaluated at a target grid point labeled by
$i$ and acts on a function sampled on source grid points $\xi_j$:
\begin{equation}
    I_i
    =
    \int_{a_i}^{b_i} d\xi\,K(\xi_i,\xi) f(\xi).
    \label{eq:app_generic_integral}
\end{equation}
Here $i$ labels the target point at which the output $I_i$ is evaluated, and
$a_i$ and $b_i$ are the integration limits for that target point.  The index
$j$ labels the source grid points used to represent $f$.  For each target point
$i$, choose quadrature nodes $\xi_{i\lambda}$ and weights $w_{i\lambda}$ on the
interval $[a_i,b_i]$, where $\lambda$ labels the quadrature nodes.  This gives
\begin{equation}
    I_i
    \simeq
    \sum_\lambda
    w_{i\lambda}K(\xi_i,\xi_{i\lambda})f(\xi_{i\lambda}).
\end{equation}
The sampled point $\xi_{i\lambda}$ generally lies off grid, so the value of
$f$ there is obtained by interpolation from the source grid:
\begin{equation}
    f(\xi_{i\lambda})
    \simeq
    \sum_j M_{i\lambda;j} f(\xi_j),
\end{equation}
where $M_{i\lambda;j}$ is the corresponding interpolation matrix.  Substituting
this relation into the quadrature sum gives
\begin{equation}
    I_i
    \simeq
    \sum_j
    \left[
    \sum_\lambda
    w_{i\lambda}K(\xi_i,\xi_{i\lambda})M_{i\lambda;j}
    \right]
    f(\xi_j)
    \equiv
    \sum_j \mathbb K_{ij}f(\xi_j).
    \label{eq:app_generic_matrix}
\end{equation}
Thus the discretized integral operator is a matrix $\mathbb K$, and an integral
iteration becomes a matrix iteration.

The one-dimensional recurrences in the flat-Earth approximation are direct
applications of this construction.  The spatial flux recurrence in
Eq.~\eqref{eq:plate_spatial_flux_recursion} and the spectrum recurrence in
Eq.~\eqref{eq:plate_spectrum_recursion} each contain a single integral operator.
After discretization, these recurrences take the schematic form
\begin{equation}
    \Phi_{i+1}
    =
    \mathbb K_{\rm plate}^{(z)}\Phi_i,
    \qquad
    \frac{dN_{i+1}}{dE}
    =
    \mathbb K_{\rm plate}^{(E)}
    \frac{dN_i}{dE}.
    \label{eq:app_plate_matrix}
\end{equation}
The entries of each matrix contain the relevant kernel, the quadrature weights,
and the interpolation coefficients.

The same idea applies when the integral operator is higher dimensional.  For
isotropic scattering in the spherical-Earth geometry, Eq.~\eqref{eq:spherical_flux_recursion}
contains the chord integral and the angular average.  Discretizing both
integrals gives
\begin{equation}
    \frac{d\Phi_{i+1}}{du}
    =
    \mathbb K_{\rm sph}
    \frac{d\Phi_i}{du}.
    \label{eq:app_spherical_flux_matrix}
\end{equation}
The spherical geometry changes the entries of $\mathbb K_{\rm sph}$ because
the sampled points lie on chords through the Earth, but it does not change the
matrix-iteration structure.

For the non-factorizable recurrence in Eq.~\eqref{eq:neumann_series_spherical}, the unknown is the full distribution on a
three-dimensional grid,
\begin{equation}
    F_{i,abc}
    \equiv
    F_i(r_a,u_b,E_c).
    \label{eq:app_full_grid}
\end{equation}
The recurrence contains the path, energy, and angular integrations in one
operator.  After flattening the multi-index $(a,b,c)$ into a single index
$\alpha$, it again has the matrix form
\begin{equation}
    F_{i+1,\alpha}
    =
    \sum_\beta
    \mathbb K_{\alpha\beta}F_{i,\beta}.
    \label{eq:app_full_matrix}
\end{equation}
The detailed entries of this matrix are implementation dependent, but the
conceptual steps are the same as in the one-dimensional construction in
Eq.~\eqref{eq:app_generic_matrix}.

\bibliographystyle{JHEP}
\bibliography{ref}

\end{document}